\documentclass[12pt]{article}
\usepackage{latexsym}
\usepackage[dvips]{epsfig}
\begin{document}

{\Large\bf Neutron Small Angle Scattering on Liquid Helium in the
Temperature Range 1.5-4.2\,K}

\vspace{1cm}
\centerline{\it Yu.M.Tsipenyuk$^1$ and
R.P.May$^2$}

\vspace{0.7cm}
$^1$ P.L.Kapitza Institute for Physical Problems RAS,  Moscow, Russia;

\qquad email: {\it tsip@kapitza.ras.ru.}

$^2$ Institut Laue-Langevin, Grenoble, France;

\qquad email: {\it roland.may@ill.fr.}

\bigskip
Subj-class: Soft Condensed Matter

\bigskip
\begin{abstract}

The small  angle neutron scattering (SANS) from liquid helium at saturated
vapour pressure in the temperature range from 1.5 to 4.2\,K was measured with
the instrument D22 of the ILL Grenoble at a wavelength of 4.6~\AA.  The zero
angle cross section is monotonically decreasing with decreasing temperature
and does not show any singularity at the lambda-point. On the other hand, we
observe a change of the slope of the temperature dependence of the second
momentum of the pair correlation function  at the lambda-point that reflects
the transition of liquid to the superfluid state.

\end{abstract}

 \section{Introduction} \label{intro}

Thermal neutrons are ideally suited to probing
the excitations in condensed-matter systems, as their wavelengths
$0.5<\lambda<10$ \AA\  are comparable to the interatomic distances, and their
energies are of the same magnitude as the excitations.
In particular, the small-angle neutron scattering (SANS)
gives valuable information on thermal fluctuations in liquids.

The only previous SANS
experiment on liquid helium has been performed by Egelstaff and London long
ago, in 1953 \cite{Egel}.  Unfortunately, there are only a few experimental
points below  the lambda point, and it is difficult to come to a definite
conclusion about the small-angle scattering behaviour in a wide temperature
region.  Modern high-flux reactors and advanced instruments have increased the
detection sensitivity by orders of magnitude.
Additionally, previous detail experiments on the static structure
factor of helium-4 have been performed only for the momentum
transferred above 1\,\AA$^{-1}$ \cite{Glyde,Crev}.

This was a motivation for us to remeasure SANS from liquid helium in a wide
temperature range.

\section{Experiment}
\label{sec:1}

The experiments were carried out at the
instrument D22  of the Institut Laue-Langevin (ILL), Grenoble, France.
We used a standard ILL Orange Cryostat modified for SANS with
quartz windows.  We filled the inner part of the cryostat, which is delimited
by transparent quartz windows at the neutron entrance and exit sides,  with
liquid helium by condensing it into the sample volume.  We then lowered the
helium temperature first by opening the cold valve, and when we came close to
the phase transition to He-II,  by directly pumping on the helium.

The experiments were performed with 4.6\,\AA\ neutrons.
The sample-to-detector distance was 2.8\,m (just avoiding shadowing on the
detector that was centred with respect to the beam).  The "collimation"
distance, i.e. the distance between the end of the neutron guide of
40\,mm (wide)  by 55\,mm (high) and the sample, was 5.6\,m.  The sample
aperture had a diameter of 10\,mm.  The scattering from the empty  cryostat at
low temperature, and the neutron and electronic noise were measured in order to
correct the data for backgrounds.  The data were radially averaged and the
container and noise contributions were subtracted by standard programs;
they were put on an absolute scale by normalizing with the
scattering from 1\,mm of water.  We also checked by measurements with a
detector setting of 14\,m that there was no central increase in the helium
scattering pattern.

\section{Results and Discussion}
\label{sec:2}

Figure 1 shows radially averaged scattered neutron
intensities  versus momentum transfer for some helium
temperatures and water normalized scattering patterns with sample background
subtracted.

The momentum transferred is  calculated on the assumption that the neutron
scattering is elastic and  is related to the full scattering
angle $\theta$ through the expression
$$|Q|={4\pi\over\lambda}\sin(\theta/2),\eqno(1)$$
where $\lambda$ is the wavelength of the neutron.

 \begin{figure}
 \centering
\epsfig{file=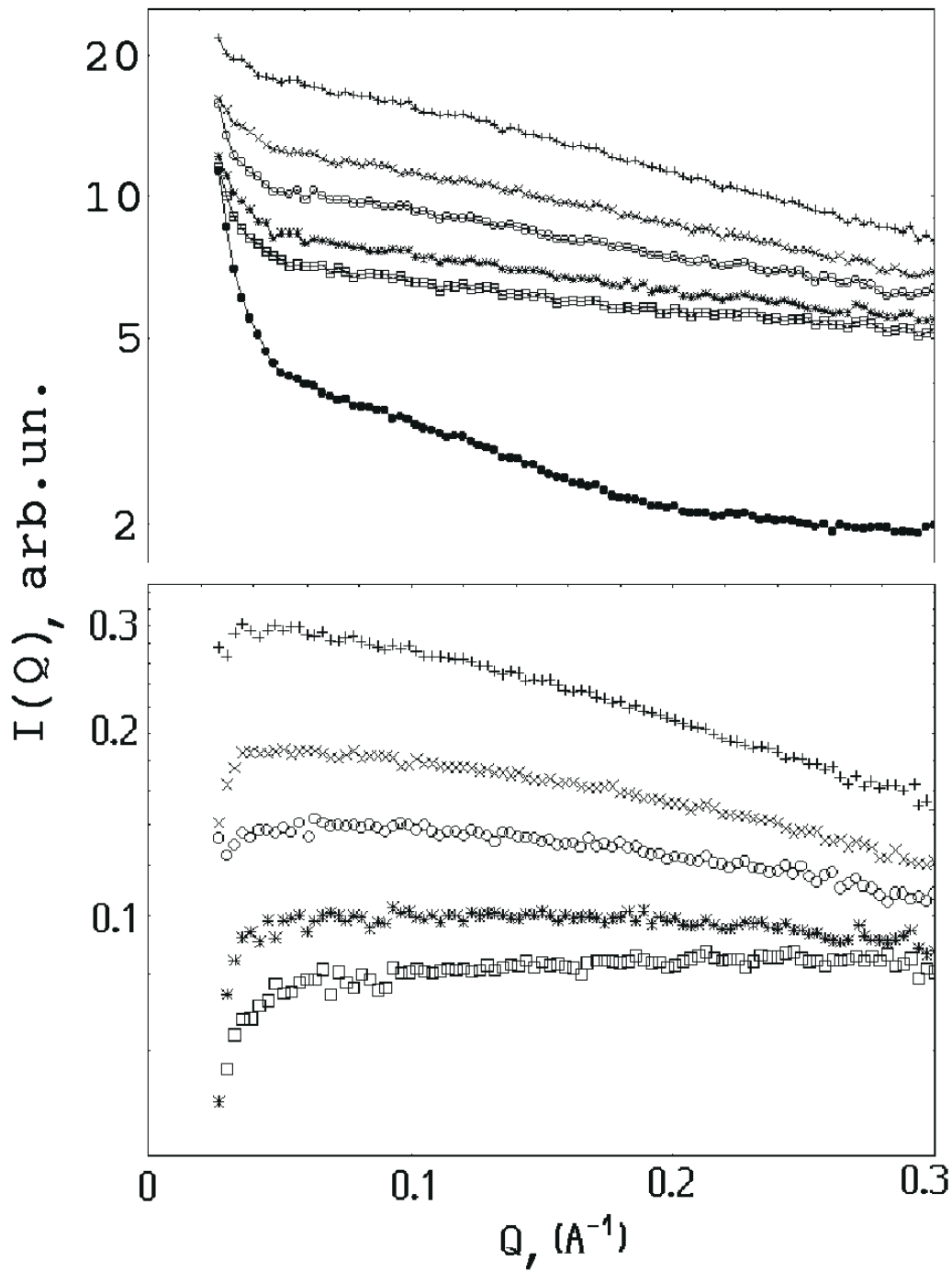,width=15cm}
\caption{ {\it Upper set of curves:}
Radially averaged intensity of scattered neutrons (in
arbitrary  units) $versus$ momentum transfer at some helium temperatures:
$+$ --- 4.09\,K, $\times$ --- 3.54\,K, $\circ$ --- 3.04\,K,  $\ast$ ---
2.18\,K, $\Box$ --- 1.54\,K, full $\Box$ --- empty cryostat. {\it Lower
set of curves}: water normalized scattering patterns with sample background
subtracted.}
\label{fig:1}
  \end{figure}

Although the background
from the cryostat is rather small in comparison with the intensity scattered
from helium, we see an increase of the background intensity
at low momentum transfer ($Q<0.03\,$\AA$^{-1}$) and, as a consequence, a
decrease of an intensuty of scattered from helium neutrons as seen in Fig.1.  A
reasonable explanation is that the small-$Q$ scattering from the cryostat is a
result of the existence of dust, cracks, scratches, that are small in size, on
the surfaces of the quartz windows.  Special experiments are needed to clarify
this situation, and that is why we analyse only experimental data in the
$Q$-range $(0.03\div0.3)$\,\AA$^{-1}$.

The essential quantity in the description of a liquid is its
structure factor $S(Q)$
that is a measure of the correlation between positions of the atoms in fluid.
The macroscopic cross section $d\Sigma/d\Omega$ of neutron scattering is
related to the static structure factor $S(Q)$ by
$${d\Sigma\over d\Omega}(Q)={\sigma_b\over4\pi}\rho_{\rm at}S(Q),\eqno(2)$$
where $\sigma_b$ is the bound  atom scattering cross section of helium (= 1.172
barn) and $\rho_{\rm at}$ is the atomic density of helium.

According to Goldstein \cite{Gold}, at finite temperature
the structure factor is given by
$$S(Q)=n_0k_{\rm B}T\chi_T+\sum\limits_{n=1}^\infty
(-1)^nQ^{2n}r_g^{2n}[(2n+1)!]^{-1},\eqno(3)$$
where $n_0$ is the number density, $k_{\rm B}$ the Boltzmann constant,
$\chi_T$ the isothermal compressibility, and  $r_g^{2n}$ is the moment of the
pair correlation function $g(r)$ defined by
$$r_g^{2n}=n_0\int r^{2n}[1-g(r)]d^3r.\eqno(4)$$

In particular, to order $Q^2$ Eq.(3) is just
$$S(Q)\simeq n_0k_BT\chi_T-r_g^2{Q^2\over 6}, \quad
 S(0)=n_0k_BT\chi_T.\eqno(5)$$

This approximation looks like the Guinier approximation
of the decay of the intensity near $Q=0$
$$I(Q)=I(0)\exp(-Q^2R_g^2/3)\simeq I(0)(1-Q^2R_g^2/3),\eqno(6)$$
where $R_g$ is the radius of gyration.

On the other hand, as it shown by  Ornstein and Zernike, in the range of
critical point the intensity of scatterred neutrons is described as
$$S(Q)={S(0)\over 1+r_c^2Q^2}\simeq
S(0)(1-r_c^2Q^2).\eqno(7)$$

We see that  there is a direct relation between $r_c$, $R_g$ and $r_g$.
The critical temperature of helium equals 5.2\,K, and thus we can definitely
say that at temperature around 4\,K the value of $R_g$, as well as $r_g$,
reflects correlation between helium atoms and we can consider them as
correlation radius. As to smaller temperatures, the physical meaning of these
quantities not so apparent.

The parabolic behaviour of the intensity of scattered neutrons at small
$Q$ is  clearly seen  in the experiment (Fig.2).

 \begin{figure}
 \centering
\epsfig{file=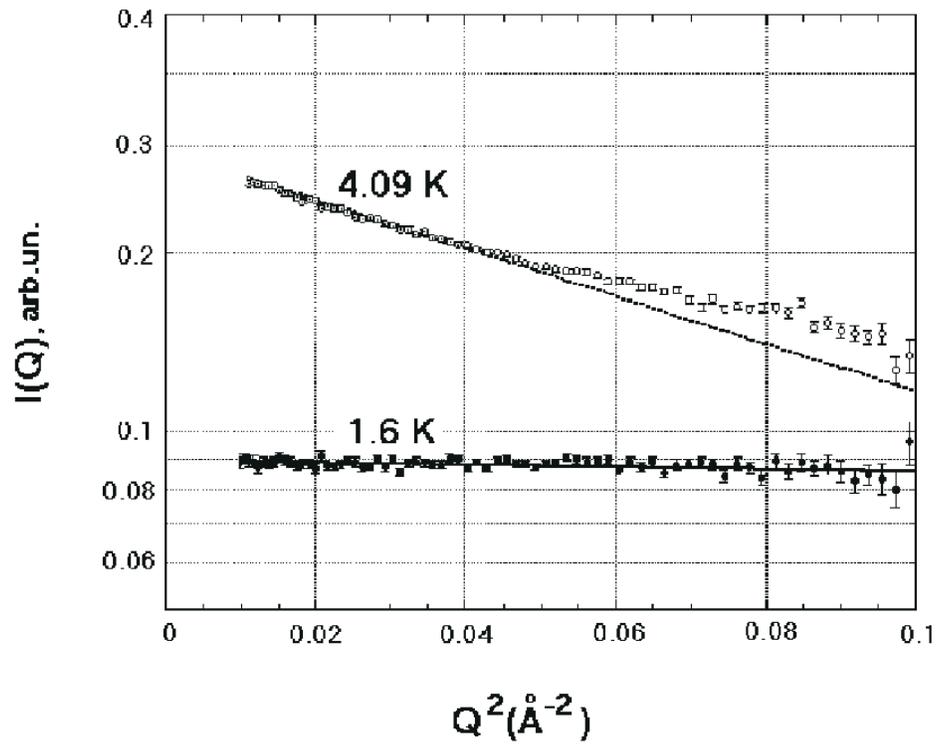,width=15cm}
\caption{
Intensity of scattered neutrons versus $Q^2$.}
\label{fig:2}
   \end{figure}

It is necessary here to note that we have not take into account
the scattering from zero-point fluctuations that leads to the linear function
of $Q$ that has been first obtained by Feynman \cite{Feyn}, but this process
takes place at lower temperatures.

We calculated the zero-angle cross section by extrapolation of the scattered
intensity to $Q$=0, and the results are shown in Fig.3.  It is
seen that the zero angle cross section is monotonically
decreasing with decreasing temperature and does not show any
peculiarity at the lambda-point.  In the same figure we show  the
theoretical zero angle cross section $S(0)$ calculated using the known
data for the temperature dependence of the helium density and the
isothermal compressibility \cite{McCarty}. At temperatures below the
lambda-point both density and isothermal compressibility are
practically independent of the temperature, and thus the
zero-angle cross section is proportional to the temperature that allows it to
fit linearly at low temperatures.
As seen, there is a temperature dependent background, but we don't
know its origin.

 \begin{figure}
 \centering
\epsfig{file=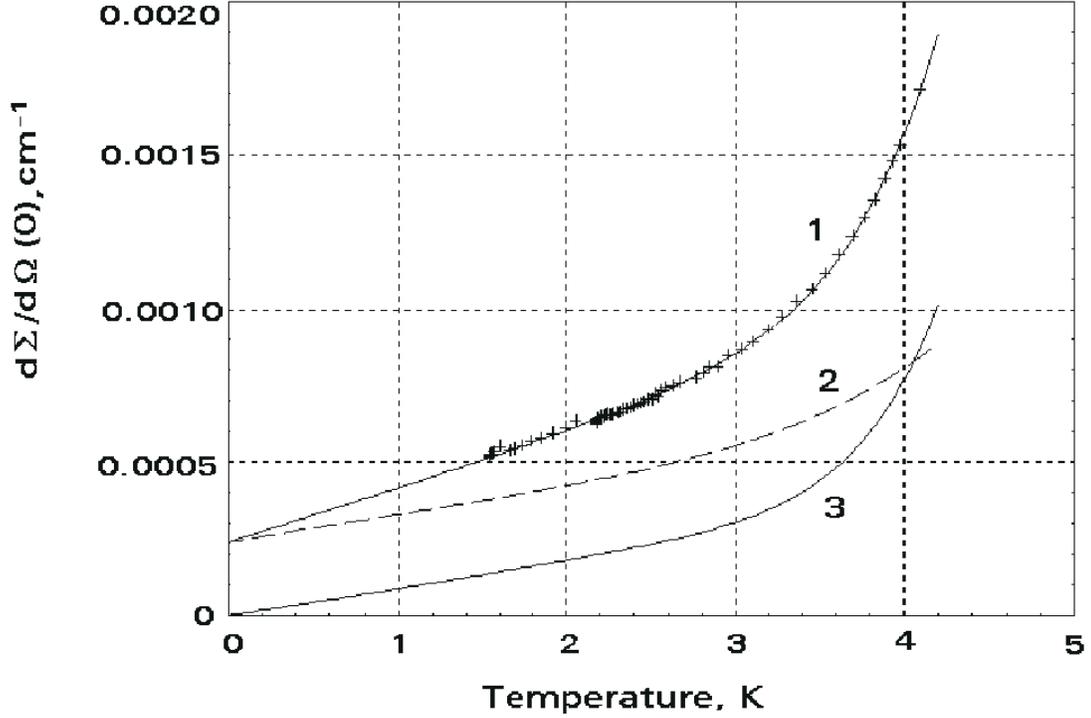,width=15cm}
\caption{
1 --- Zero-angle cross sections ($+$), solid line ---  its fit with
linear $Q$-dependence at low temperatures;  2 --- fit of measured background;
3 --- theoretical zero-angle cross section.} \label{fig:3}
 \end{figure}

We have measured SANS with a temperature interval 0.1-0.2\,K that has provided
the first comprehensive data on the correlation radius in the normal and
superfluid state of helium. The results are shown in Fig. 4.
The presented in Fig.1 data were corrected for constant background that
corresponds to $d\Sigma/d\Omega(0)$ at zero temperature. It is seen that the
temperature dependence of the correlation radius behaves like for a normal
liquid, i.e.  it decreases with decreasing temperature.  However, at the
superfluid transition temperature (2.2\,K) we see a pronounced change of its
slope.

 \begin{figure}
\centering
\epsfig{file=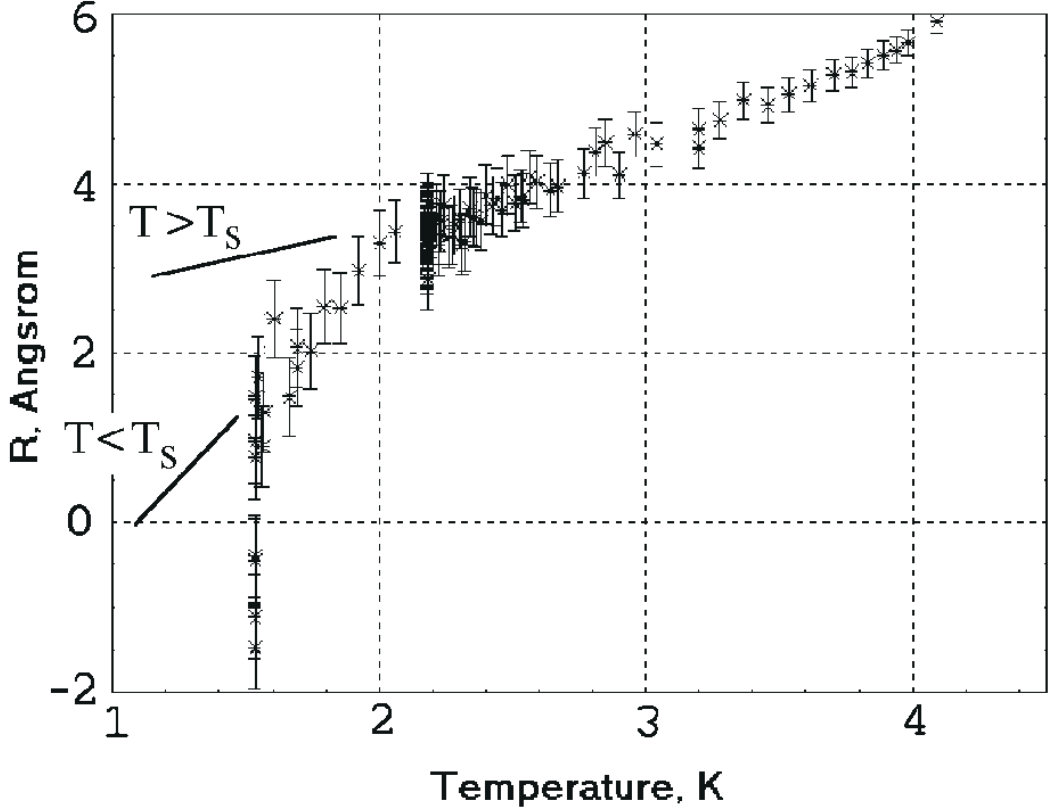,width=15cm}
\caption{
Radius of gyration $R$ as a function of temperature. The
negative sign of the radius means the change of the derivative of the
function $I(Q^2)$. For clarity the slope of  $R(T)$ in the vicinity of the
temperature  of the superfluid transition $T_s$ is shown by solid
lines.} \label{fig:4}
  \end{figure}

At a qualitative level the observed loss of spatial correlations in
liquid helium is due to a rearrangement of atoms to create the
necessary spaces for delocalisation to occur. Nevertheless, this intuitive
explanation has not theoretical basis. For instance, using Monte Carlo method
Mayers \cite{Mayers} argued that the link between spatial correlations and
Bose-Einstein condensate fraction is a geometrical consequence of the
hard-core repulsion between atoms.  The excitation of rotons has also been
proposed  by Masserini {\it et al.} \cite{Masser} as an alternative
explanation for the increase in spatial correlation as the temperature is
rased in the superfluid state.

In this connection, we have to emphasize that in our SANS experiment the
distance scale probed is greater than interatomic distances, and the
correlation length obtained characterizes long-range order in liquid helium
whereas in  all previous experimental and theoretical works the behaviour of
the first peak in the $S(Q)$ was considered, i.e. short-range correlations at
$Q$-value about 2 angstrom$^{-1}$. In our SANS experiment we deal with
$Q<0.3\,$\AA$^{-1}$.  As opposed to previous experiments  we have studied
long-range correlation between helium atoms.

\section{Acknoledgements}
\label{sec:3}

The authors are  deeply grateful
A.S.Stepanov  and S.M.Apenko for fruitful discussions of this work.

\end{document}